**Diffusion-Controlled Anion Conversion into Dense Polycrystalline and Single-Crystalline Oxyhydrides**

Masaya Fujioka[1,2]*, Mihiro Hoshino[2], Suguru Iwasaki[3], Katsuhiro Nomura[1], Aman Sharma[1], Mineyuki Hattori[4], Zhongxu Hu[5], Takayoshi Katase[5], Reina Utsumi[6], Yuki Nakahira[6], Hiroyuki Saitoh[6],

[1] Innovative Functional Materials Research Institute, National Institute of Advanced Industrial Science and Technology (AIST), 4-205 Sakurazaka, Moriyama-ku, Nagoya, Aichi 463-8560, Japan

[2] Research Institute for Electronic Science, Hokkaido University, Kita 20, Nishi10, Kita-ku, Sapporo, Hokkaido 001-0020, Japan

[3] Department of Industrial Chemistry, Faculty of Engineering, Tokyo University of Science, Katsushika, Tokyo, 125-8585, Japan

[4] Research Institute for Material and Chemical Measurement, National Metrology Institute of Japan (NMIJ), National Institute of Advanced Industrial Science and Technology (AIST), Tsukuba, Ibaraki 305-8565, Japan

[5] MDX Research Center for Element Strategy, Institute of Integrated Research, Institute of Science Tokyo, 4259 Nagatsuta, Midori, Yokohama 226-8501, Japan

[6] National Institutes for Quantum Science and Technology (QST), 1-1-1 Kouto, Sayo, Hyogo 679-5148, Japan

* E-mail: m.fujioka@aist.go.jp

Abstract

Oxyhydrides represent a new class of functional materials, yet the synthesis of dense polycrystals or single-crystals suitable for transport studies remains a significant challenge due to hydrogen desorption at elevated temperatures. The co-diffusion of oxygen and hydrogen in densely sintered $BaTiO_3$ enables the topochemical formation of millimeter-scale bulk $BaTiO_{3-x}H_x$ via high-pressure diffusion control (HPDC). Hydride ions selectively occupy oxygen-deficient sites, as confirmed by neutron diffraction, TPD, TG, and NMR. Systematic tuning of the hydrogen content and precise control of the electronic conductivity were achieved via HPDC. Hydrogen desorption analysis reveals distinct bonding states between near-surface and interior-bulk regions, which significantly affect the oxynitride conversion under $N_2$ flow. Importantly, the diffusion-based nature of HPDC allows direct anion conversion even in single-crystalline oxides, as demonstrated by the synthesis of $SrTiO_{3-x}H_x$ single crystals. These results establish HPDC as a general platform for accessing dense, metastable oxyhydrides with tunable anionic composition and transport properties.

## 1. Introduction

Engineering the anion sublattice in oxides offers a powerful lever to tailor their physical, chemical, and structural properties, thereby unlocking novel functionalities[1]. In particular, perovskite oxides allow for the substitution of oxygen anions with hydride ions, and once such hydride incorporation is achieved, further conversion into oxynitrides can be realized by heat treatment under nitrogen or ammonia atmospheres[2]. These tunability of anion species with different valence states: $H^-$ or $N^{3-}$ enables further modification of electron, ion, and phonon transport behaviors. However, probing these intrinsic properties requires densely sintered oxyhydrides with strong intergranular connectivity. This presents a central conflict: hydrogen begins to escape at relatively low temperatures around 400–500 °C[3], yet forming dense microstructures requires heating beyond the stability range of oxyhydride phases [4]. Resolving this trade-off remains a key challenge in realizing the transport functionality of oxyhydrides.

Diffusion control provides a viable strategy to resolve this trade-off by enabling anion conversion after dense sintering. Specifically, it allows the formation of bulk oxyhydrides while preserving the original morphology, through the simultaneous diffusion-driven formation of oxygen vacancies and hydride incorporation. Such selective diffusion of specific elements to modulate chemical composition within solid materials has already been demonstrated in several systems.

For example, in silicon clathrates or boron cluster compounds, weakly bonded atoms in covalent frameworks were homogeneously extracted by intentionally introducing chemical potential gradients[5]. This elemental extraction technique was further applied under high pressure, as high-pressure diffusion control (HPDC), to overcome diffusion limitations associated with volume contraction[6]. In layered and fibrous materials with van der Waals gaps, intercalation provides an effective means of tuning chemical composition without significantly disrupting the original crystal structure. Notable examples include monovalent cation-assisted multivalent intercalation[7] and proton-driven ion introduction via corona discharge[8]. In ion-conducting glass based on phosphate frameworks, cations within the structure become mobile at temperatures above the glass transition point, enabling nearly 90% of them to be replaced with protons under an applied voltage in a hydrogen atmosphere[9].

All these systems share a key feature: at specific temperatures, certain atomic species can migrate while the structural framework remains largely intact. In such systems, carefully designed driving forces, such as chemical potential gradients and electric fields, can induce anisotropic diffusion of selected atomic species[10]. Such compositional transformations are typically described as topotactic reactions[11]. A distinguishing advantage of diffusion control lies in its ability to drive topotactic transformations deep into dense bulk solids. By providing an appropriate environment for anisotropic diffusion, compositional modulation can be achieved throughout the entire sample, even in densely sintered polycrystals or millimeter-scale single crystals[5a], thereby enabling direct access to the intrinsic electronic, ionic, and thermal transport properties of the resulting phases.

Building upon prior developments in diffusion control, this study extends the method to oxide materials, one of the most widely studied inorganic material families. In oxides with corner-sharing polyhedral units, oxide-ion diffusion tends to occur to some extent at elevated temperatures. Moreover, numerous mechanisms have been explored to further enhance this diffusion behavior.[12] Moreover, various mechanisms have been explored to further enhance this diffusion behavior. [13] These features of oxides can thus be leveraged as a basis for diffusion-mediated anion conversion, enabling the bulk-phase transformation into oxyhydrides from oxides, and facilitating direct evaluation of their transport properties. This study demonstrates anion conversion in densely sintered bulk $BaTiO_3$ and bulk single-crystalline $SrTiO_3$ to oxyhydrides via the HPDC method.

## 2. Results and Discussion

### 2.1 Conversion from oxide to oxyhydride and characterization

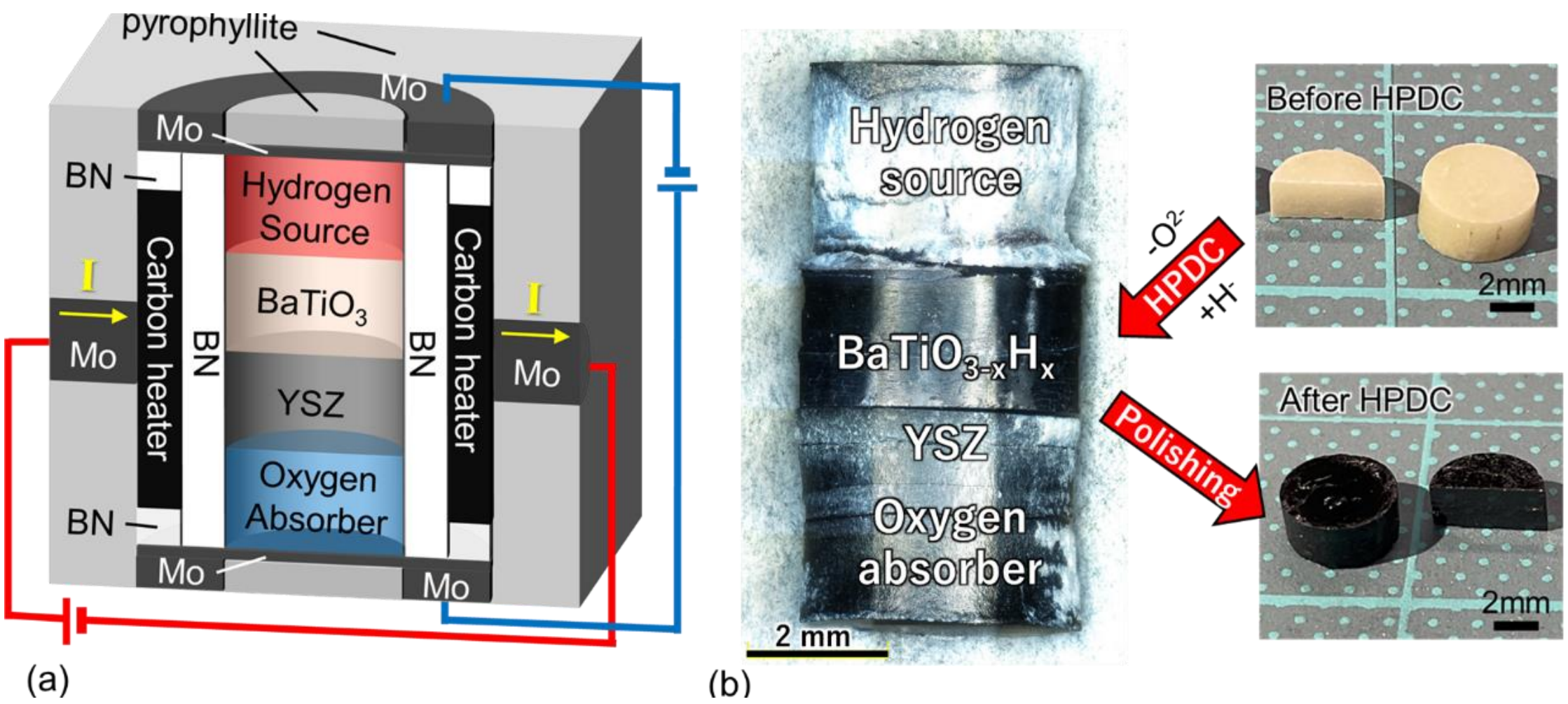

Figure 1 | High-pressure diffusion control (HPDC) setup and bulk oxyhydride formation

(a) Schematic of the HPDC cell enabling simultaneous application of heat, pressure, and electric field for topotactic anion conversion. (b) Optical images of each component after HPDC and cross-sectional view of the bulk $BaTiO_3$ sample before and after treatment, showing uniform color change indicative of hydrogen incorporation throughout the pellet.

Figure 1(a) illustrates a schematic cross-section of an HPDC cell. The cell, modified from a cubic-type high-pressure cell, enables the simultaneous control of temperature, pressure, and electric field. Temperature is regulated through lateral molybdenum (Mo) electrodes, while voltage can be independently applied across the sample space using vertical Mo electrodes. The applied pressure ensures intimate chemical and physical contact both within and between materials, while also accommodating gradual volume changes induced by elemental migration [6b].

An optical image of the inside of sample spaces after HPDC, and a cross-section of samples before and after HPDC treatment are shown in Figure 1(b), clearly indicating that the originally white $BaTiO_3$ has undergone a uniform color change to blue-black throughout the interior. Although multiple parameters, such as pressure, temperature, applied voltage, the composition of the hydrogen source and oxygen absorber within the sample space (in this study, $MgH_2$ or $CaH_2$ for hydrogen supply and Ti for oxygen absorption), and the thicknesses of each component, must be carefully controlled for successful oxide-to-oxyhydride conversion via HPDC, our study reveals that three parameters are particularly critical in determining the extent of conversion: the sintering temperature of the precursor $BaTiO_3$, the HPDC treatment temperature, and the ratio of Ti to yttria-stabilized zirconia (YSZ) in the oxygen absorber. Details of these parameters are provided in the Methods section.

Structural analyses of the HPDC-treated samples were performed by Rietveld refinements of powder neutron diffraction using Z-code[14]. It is known that the tetragonal structure of $BaTiO_3$ approaches the cubic symmetry upon hydrogen substitution[15], and a similar structural evolution was observed in the HPDC-treated samples. Although the difference was marginal, the refinement assuming a tetragonal model yielded better convergence (Figure S1, Table S1, and S2); hence, the tetragonal symmetry was adopted in this study. Figure 2(a) shows the Rietveld refinement assuming hydrogen occupancy at oxygen-deficient sites, represented as $BaTiO_{3-x-y}H_x\square_y$, while Figure 2(b) shows the result based on a model with oxygen vacancies only, represented as $BaTiO_{3-y}\square_y$. Both models exhibited comparable refinement with no impurity phases. In the hydrogen-substituted model, independent refinement of hydrogen and oxygen occupancies converged reliably, yielding $x = 0.214$ and $y = 0.001$. These values indicate that most oxygen-deficient sites were filled by hydride ions, in good agreement with the hydrogen content $x = 0.215$, estimated from temperature-programmed desorption (TPD; Figure 2(c)) and the total oxygen deficiency $x + y \approx 0.21$, estimated from Thermogravimetry (TG; Figure 2(d)).

In contrast, the model assuming only oxygen vacancies yielded y = 0.354, which significantly deviates from the TPD and TG results. Notably, similar refinement results were obtained for a sample with lower hydrogen content (Figure S1, Table S1, and S2). These results strongly support the hydrogen-substituted model and indicate that the $BaTiO_3$ pellet was successfully converted to a $BaTiO_{3-x}H_x$ pellet through diffusion, with most of the oxygen vacancies being effectively replaced by hydride ions.

In diffusion-based synthesis, where oxide anions are replaced with hydride ions, it is important to assess whether other elements from the hydride source or the oxygen absorber unintentionally diffuse into the $BaTiO_3$ bulk. Figure 2(e) presents EDS results obtained from the polished surface of the sample after HPDC treatment. No detectable signals of Mg from the $MgH_2$ in the hydride source or Zr and Y from the YSZ were observed on either side of the sample. Furthermore, assuming a Ti/Ba ratio of 1 in the pristine $BaTiO_3$, no significant deviation in the elemental ratio was found after HPDC treatment, indicating that undesired elemental diffusion did not occur.

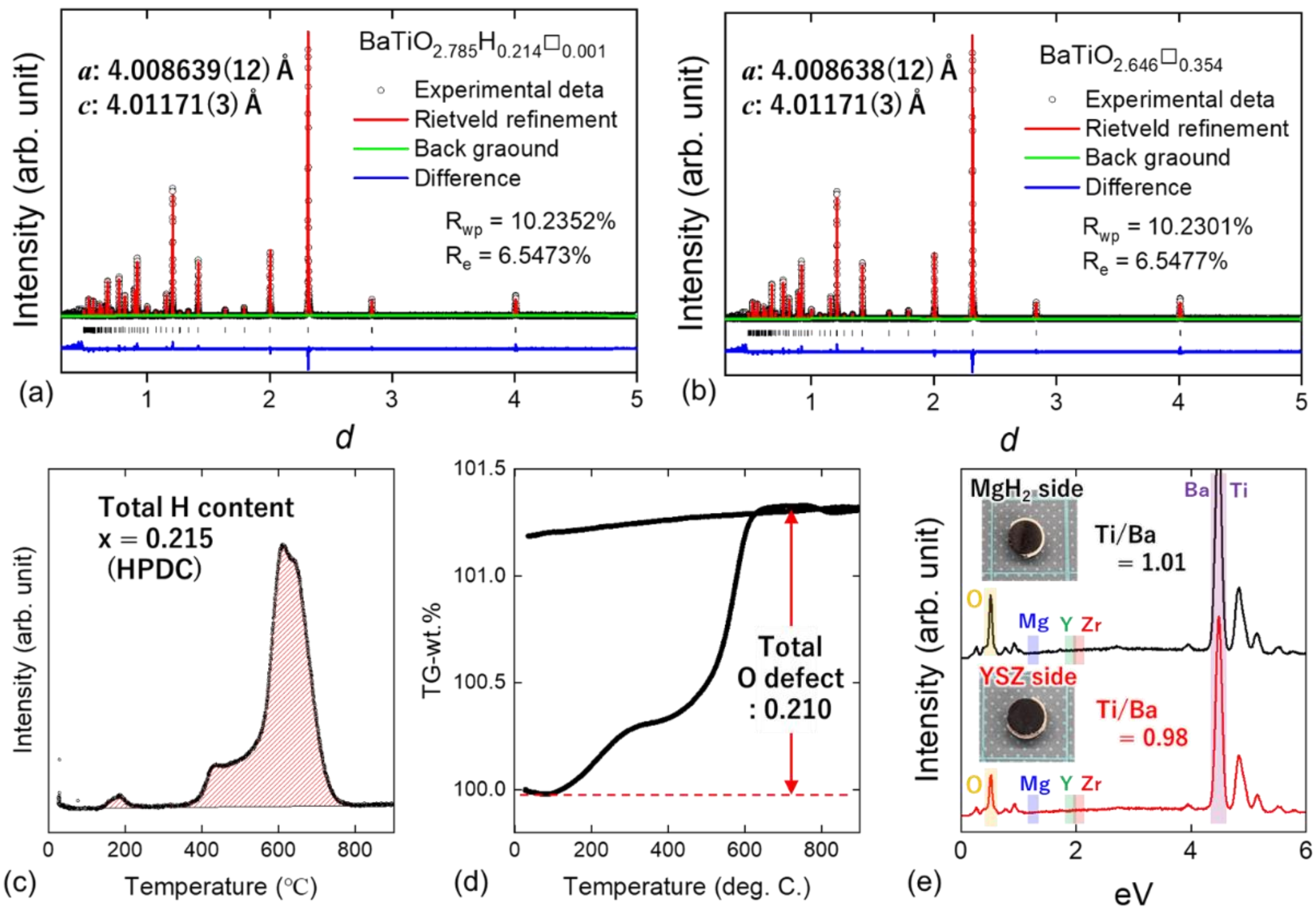

Figure 2. Structural characterization and compositional analysis of HPDC-treated $BaTiO_{3-x}H_x$.

(a, b) Rietveld refinements of neutron diffraction patterns assuming hydrogen incorporation (a) and without hydrogen (b). (c) Hydrogen content estimated by TPD. (d)

Total oxygen deficiency determined from TG analysis. (e) EDS analysis after surface polishing, showing no detectable diffusion of Mg, Y, or Zr and confirming composition retention after HPDC.

### 2.2 Hydrogen state and size effect on $BaTiO_{3-x}H_x$

Nuclear magnetic resonance (NMR) measurements were conducted to investigate the hydrogen state in $BaTiO_{3-x}H_x$ obtained via HPDC, as shown in Figure 3(a). A broad signal corresponding to hydride was observed at −25 ppm, consistent with the profiles reported for previously studied $BaTiO_{3-x}H_x$[16]. A negative chemical shift reflects increased electron density around hydrogen atoms, indicating the formation of hydride ions ($H^-$). In addition, a positive peak was also observed, which is likely attributable to surface hydroxyl groups (OH), as suggested in earlier studies[16].

To further investigate the binding environments of hydrogen, TPD measurements were carried out, which revealed that the initial sintering degree and grain sizes of the parent $BaTiO_3$ are critically important for the oxyhydride conversion via HPDC. Indeed, it is well established that sintering temperature significantly influences the microstructure and functional properties of $BaTiO_3$. As the sintering temperature increases beyond approximately 1250 °C, $BaTiO_3$ undergoes rapid grain growth, resulting in crystallite sizes on the order of tens of microns[17]. Even within $BaTiO_3$ samples of identical composition, differences in processing conditions and grain size can lead to variations in structural factors, such as domain configuration and point defect distribution, not only between grain interiors and surfaces, but also among individual grains[17-18]. These structural variations critically influence the functional properties of $BaTiO_3$, including oxygen diffusion behavior at elevated temperatures.

TPD measurements were performed on finely ground powders of three samples to examine their hydrogen release characteristics, as shown in Figure 2(b). All samples were subjected to identical HPDC conditions but were prepared from parent $BaTiO_3$ sintered at 1100, 1200, and 1300 °C, respectively. As the sintering temperature increased, the relative density of the $BaTiO_3$ pellets significantly improved, reaching 76% (4.57 g/cm$^3$), 89% (5.37 g/cm$^3$), and 96% (5.79 g/cm$^3$), respectively. The 1100 °C sample exhibited a single hydrogen desorption peak around 450 °C, similar to the previous reports. In contrast, the 1200 °C sample showed two distinct peaks at approximately 450 °C and 650 °C, suggesting the presence of hydrogen in two different bonding states. When the sintering temperature of parent $BaTiO_3$ was further increased to 1300 °C, the 650 °C peak became significantly more prominent. As a result, a clear trend was observed: higher densification of the parent $BaTiO_3$ led to increased hydrogen incorporation. This behavior can be attributed to the diffusion-based nature of the HPDC process: in samples with lower densification and higher lattice strain at grain boundaries, oxygen diffusion across them is suppressed, hindering the formation of oxygen-deficient sites required for hydride substitution. Notably, even in the low-density $BaTiO_3$ sintered at 1100 °C, the hydrogen content could be increased up to x = 0.195 by raising the HPDC treatment temperature from 650 °C to 900 °C, thereby activating oxygen diffusion at higher temperatures (Figure S2).

As described above, hydrogen desorption around 450 °C is also observed in conventionally synthesized $BaTiO_{3-x}H_x$, typically prepared by co-sintering $BaTiO_3$ powder (~0.1 μm) with $CaH_2$[19]. In contrast, hydrogen desorption at 650 °C has not been reported previously and appears to be a unique feature of the HPDC-treated samples. To investigate the origin of this difference, three samples with different particle sizes were prepared: one synthesized by the conventional method using $BaTiO_3$ nanopowder (~0.1 μm), and the remaing two derived from the same HPDC-treated bulk pellet prepared from $BaTiO_3$ sintered at 1300 °C: one coarsely crushed (~100 μm) and the other finely ground (~1 μm) using a mortar and pestle, as shown in Figure 3(c). In addition, optical images of the three samples are shown in Figure S3(a), exhibiting that the color becomes progressively darker with increasing particle size, even though their hydride ion contents are nearly identical.

The nanopowder exhibited a hydrogen desorption peak around 450 °C (Figure 3(d)). In contrast, the coarse powder released most of its hydrogen near 650 °C. Although the fine powder was obtained simply by further grinding the coarse powder, it exhibited three distinct desorption peaks at approximately 150 °C, 450 °C, and 650 °C. The mechanical grinding process induced the additional hydrogen desorption observed at 150 °C and 450 °C. In particular, the 150 °C peak likely corresponds to surface hydrogen that is weakly adsorbed as a result of mechanical damage. Notably, the desorption profile exhibits a clear splitting of peaks between 450 °C and 650 °C. This suggests a significant difference in the bonding states of hydrogen between the near-surface and the bulk interior, with the higher-temperature peak at 650 °C indicating a more strongly bound state, as illustrated in Figure 3(c), rather than a behavior governed solely by diffusion-limited kinetics. Indeed, reducing the heating rate from 400 °C/h to 100 °C/h led to more pronounced peak splitting, as shown in Figure S3(b).

Whereas conventional solid-state reactions proceed by reducing particle size and increasing interfacial contact, HPDC drives anion conversion by promoting grain growth and lattice ordering in the parent phase, thereby enhancing oxygen diffusivity. These results highlight the unique capability of HPDC to control anisotropic, macroscopic diffusion across bulk materials.

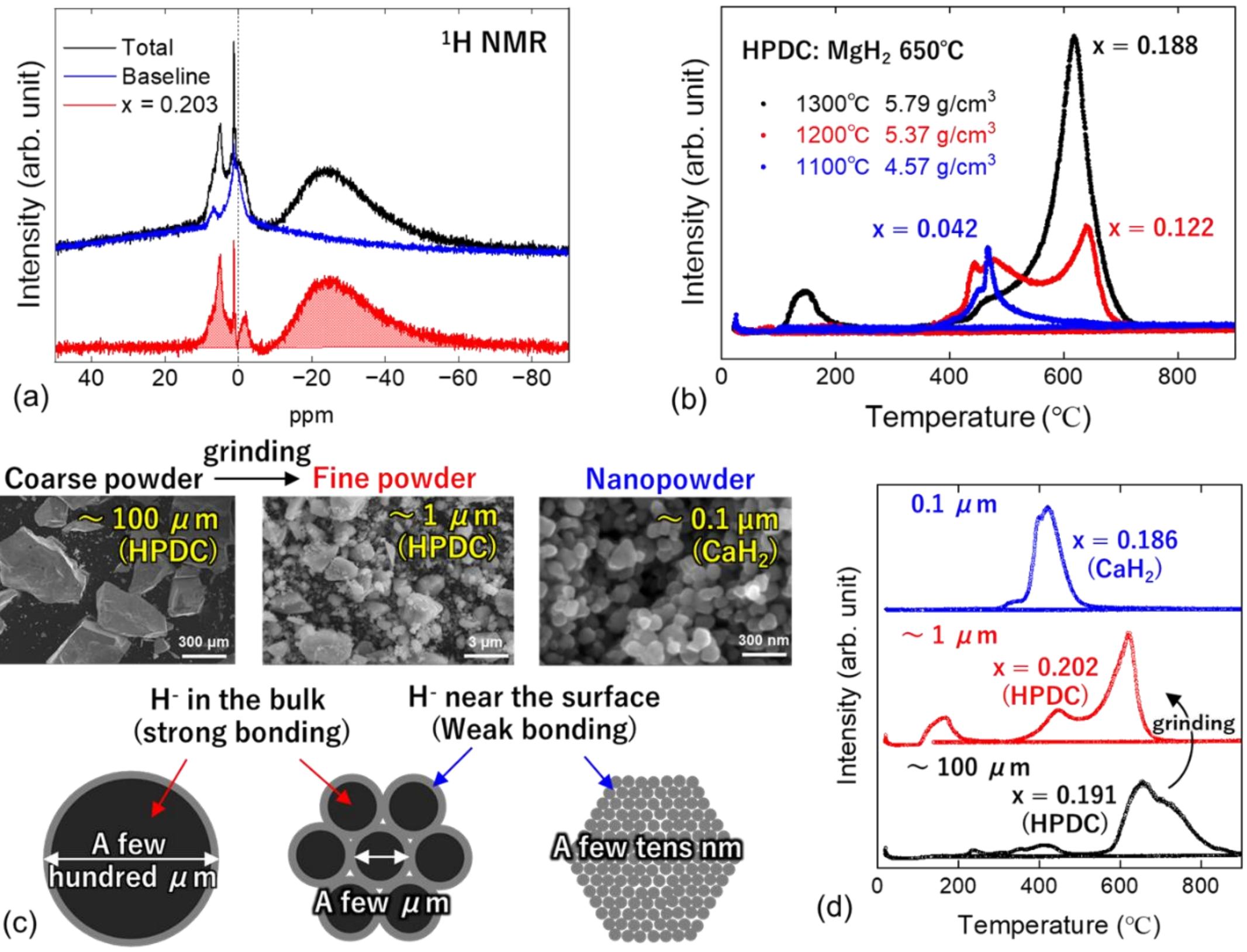

Figure 3. Hydrogen states and size effect on $BaTiO_{3-x}H_x$.

(a) Solid-state ¹H NMR spectrum for $BaTiO_{3-x}H_x$ (x = 0.203). (b) Hydrogen desorption profiles from TPD measurements for HPDC-treated samples prepared using parent $BaTiO_3$ sintered at 1100, 1200, and 1300 °C. (c) SEM images for each sample and schematic illustration of size-dependent hydrogen bonding states in $BaTiO_{3-x}H_x$. (d) TPD profiles for samples with different particle sizes for the $CaH_2$-treated nanopowder (~100 nm) and HPDC-treated coarse (~100 μm) and fine (~1 μm) powders.

### 2.3 Electrical transport properties of bulk $BaTiO_{3-x}H_x$

A major advantage of HPDC is that it enables anion conversion into oxyhydrides without altering the original sintered morphology, thereby yielding a highly dense bulk material. This, in turn, allows for the evaluation of the intrinsic transport properties of $BaTiO_{3-x}H_x$. Figure 4(a) shows the temperature dependence of electrical resistivity for bulk $BaTiO_{3-x}H_x$ samples with various hydrogen concentrations ranging from 0.032 to 0.215, as quantified by TPD in Figure 4(b). The hydride content systematically increased with increasing Ti content in the oxygen absorber, as described in the Methods section.

Although diffusion generally proceeds in response to an electrochemical potential gradient, prolonged HPDC treatment for 40 hours results in a nearly uniform hydrogen distribution throughout the sample. As shown in Figure S4, the hydrogen concentration on the oxygen absorber side is approximately 10% higher than that on the hydrogen source side. Throughout this study, such a slight increase in hydrogen content was consistently observed on the oxygen absorber side, suggesting a conversion mechanism in which oxygen vacancies are initially created and subsequently filled by hydride ions.

The obtained samples were cut into bars with dimensions of 4.2 x 1.2 x 0.5 $mm^3$. The resistivity slightly increased at lower temperatures for the samples with x = 0.036, 0.084, and 0.088. However, as the hydrogen concentration increases, the electrical resistivity decreases linearly with temperature, indicating metallic behavior. At x = 0.219, the resistivity at 300 K reached $1.6 \times 10^{-3}$ Ωcm. These conduction properties are comparable to those of reported epitaxial thin films $BaTiO_{3-x}H_x$ with x = 0.3 ($0.8 \times 10^{-3}$ Ωcm)[20], demonstrating that HPDC enables the synthesis of dense bulk oxyhydrides with significantly reduced intergranular resistance.

These findings underscore the potential of the HPDC method as a versatile platform for investigating the intrinsic electronic, ionic, and thermal transport properties of dense bulk oxyhydrides with tunable hydrogen content, which have been difficult to evaluate using conventional synthesis approaches.

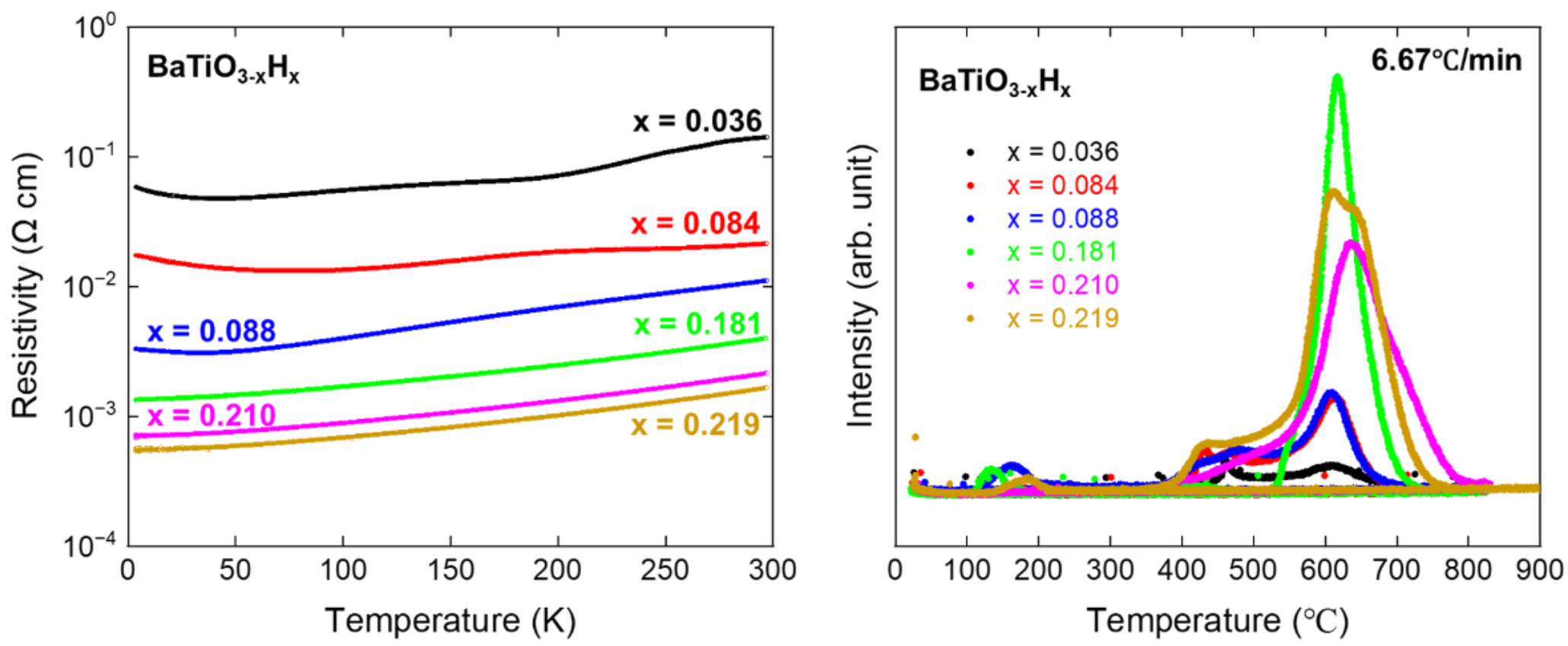

Figure 4. Electrical transport in dense $BaTiO_{3-x}H_x$ bulk samples.

(a) Temperature dependence of resistivity for samples with various hydrogen contents (x). (b) Corresponding hydrogen desorption profiles measured by TPD. The estimated x are indicated for each sample.

### 2.4 Anion conversion from oxyhydride to oxynitride via HPDC treatment

One of the key features of oxyhydrides is their ability to undergo anion conversion from hydride ($H^-$) to nitride ($N^{3-}$). Figure 5(a) presents the XRD patterns and color changes of the samples after heat treatment under a nitrogen atmosphere at various temperatures. At 800 °C and 900 °C, the formation of $Ba_2TiO_4$ was observed, while no other impurity phases were detected. In contrast, samples treated at 700 °C and 600 °C exhibited a green coloration. Similar color changes have been previously reported upon nitrogen substitution, indicating a narrowing of the band gap[2].

Figure 5(b) shows the difference in hydrogen desorption behavior under argon and nitrogen atmospheres for two $BaTiO_{3-x}H_x$ samples: one prepared by co-sintering $BaTiO_3$ nanopowder with $CaH_2$, and the other obtained by HPDC treatment of $BaTiO_3$ sintered at 1300 °C. In both samples, the hydrogen desorption temperature was found to be higher under an argon atmosphere compared to nitrogen. This effect was particularly pronounced in the HPDC-treated sample, where most of the hydride ions were strongly bound within the interior of the crystallites. Furthermore, as shown in Figure 5(c), the HPDC-treated sample was held at 500 °C while the atmosphere was switched from Ar to $N_2$. A sharp increase in hydrogen desorption was observed immediately after the introduction of $N_2$.

These results indicate that nitrogen is essential for promoting hydrogen release from the particle interior, suggesting a mechanism in which hydride ions are retained until nitrogen atoms reach the hydride sites deep inside the crystallites, thereby preventing the formation of oxygen vacancies. In contrast, the hydrogen near the surface can be released without nitrogen substitution. As shown in Figure 5(d), although both the HPDC- and $CaH_2$-treated samples contained nearly the same amount of hydrogen before nitrogen treatment, the amount of nitrogen ultimately incorporated was significantly higher in the HPDC sample. This difference was clearly observed even within the sensitivity range of EDS analysis, based on the average of five independent measurements. These findings demonstrate that HPDC-treated samples more readily undergo oxyhydride-to-oxynitride conversion even under atmospheric-pressure nitrogen flow conditions.

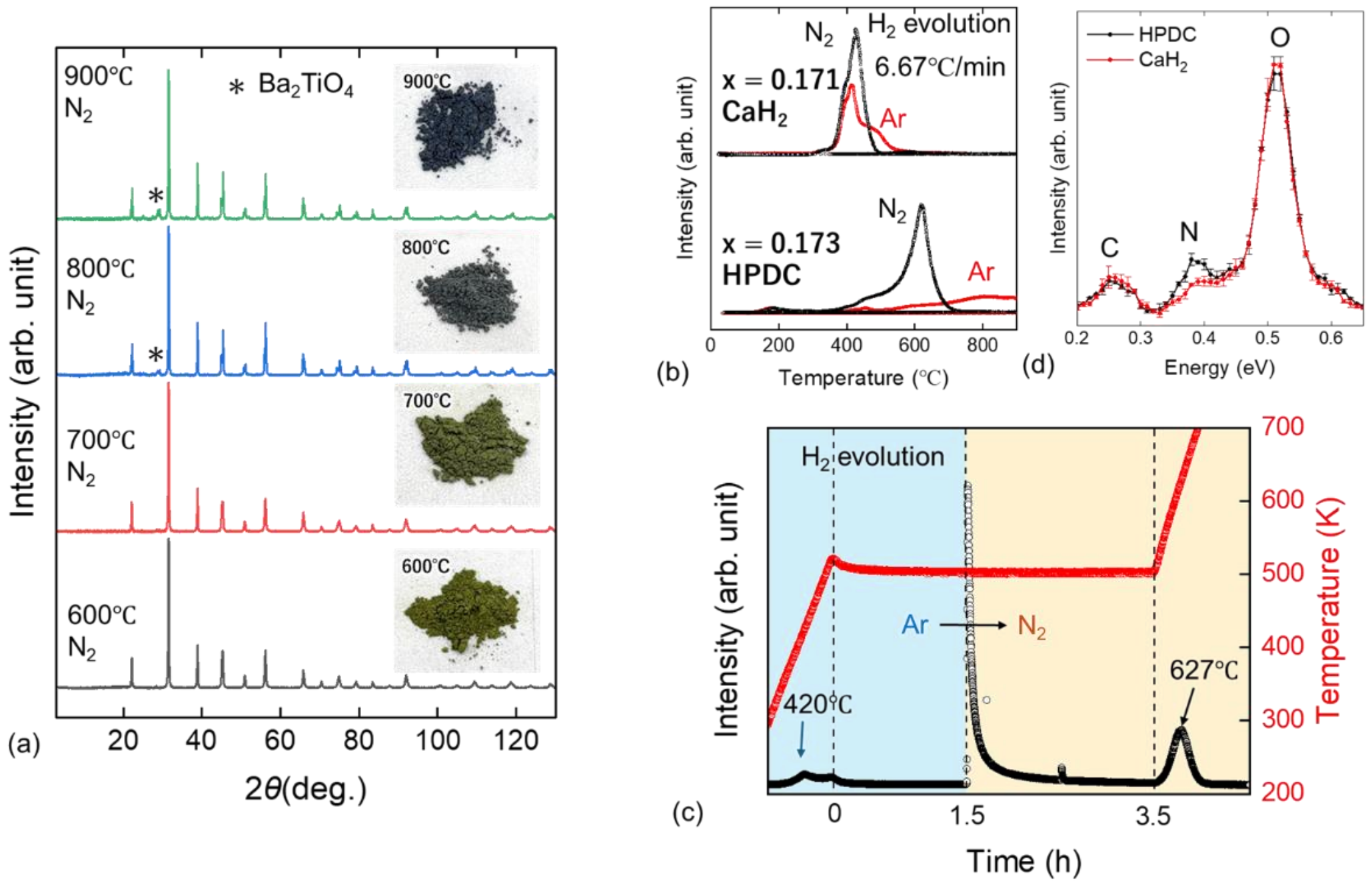

Figure 5. Anion exchange from hydride to nitride in $BaTiO_{3-x}H_x$.

(a) XRD and optical color changes after $N_2$ annealing at different temperatures. (b) Hydrogen desorption profiles under $N_2$ and Ar atmospheres for HPDC- and $CaH_2$-treated samples. (c) Accelerated hydrogen release upon switching the atmosphere from Ar to $N_2$. (d) Nitrogen EDS spectra for the $N_2$-annealed samples shown in (b), averaged over five independent measurements.

## 2.5 Further possibility of HPDC: single-crystalline oxyhydrides

In contrast to conventional solid-state reactions, which rely on intimate mixing of raw powders and local interdiffusion, HPDC proceeds more effectively when the parent oxide exhibits higher densification and larger grain growth (Figure 3). It is well established that grain boundaries act as significant scattering centers for oxide-ion diffusion[21]. Therefore, their elimination through high-temperature sintering significantly enhances oxide-ion mobility and facilitates anion conversion during HPDC. This mechanism, based on macroscopic and anisotropic diffusion across the sample space, is expected to be applicable not only to densely sintered polycrystals but also to bulk single-crystalline materials.

In general, the growth of oxide single crystals typically relies on high-temperature melt processes, where slow crystallization from the molten phase is essential[22]. This requirement renders the synthesis of single-crystalline oxyhydrides inherently difficult, as these materials tend to release hydrogen at much lower temperatures. Thus, the HPDC method opens a new pathway toward single-crystalline oxyhydrides, previously considered inaccessible due to hydrogen loss at elevated temperatures.

Figure 6(a) presents optical images of both single-crystalline and polycrystalline $SrTiO_3$ before and after HPDC treatment, while Figures 6(b) and 6(c) show the corresponding XRD patterns. When single-crystalline $SrTiO_3$ was adopted as the parent oxide, the highly oriented structure remains intact, although a decrease in diffraction intensity was observed. As shown in Figure S5, TPD analysis reveals hydrogen release over a broad temperature range from ~400 °C to ~700 °C, with the hydrogen content quantified as $x = 0.123$ for the single-crystalline sample and $x = 0.111$ for the polycrystalline counterpart. These findings demonstrate that HPDC enables the formation of single-crystalline oxyhydrides via a diffusion-based anion conversion pathway. The emergence of single-crystalline oxyhydrides in a metastable state unlocks new opportunities for investigations into their fundamental structure and functional potential.

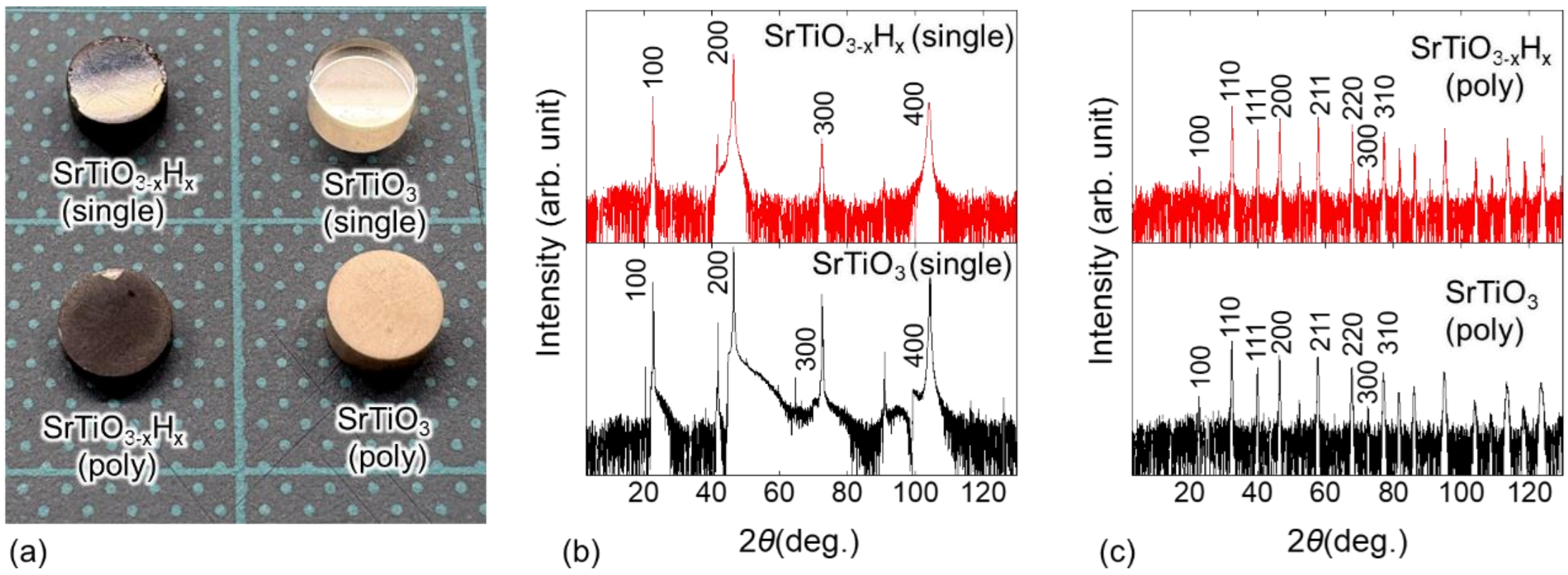

Figure 6. Formation of single-crystalline oxyhydride via HPDC.

(a) Optical images of single- and polycrystalline $SrTiO_3$ before and after HPDC. (b, c) XRD patterns demonstrating the retention of crystallinity after hydride-ion conversion for (b) single-crystalline and (c) polycrystalline samples.

## 3. Conclusion

This study demonstrates that high-pressure diffusion control (HPDC) offers a powerful and versatile strategy for synthesizing metastable oxyhydrides from oxide precursors while retaining the original sintered morphology. Neutron diffraction, TPD, TG, and NMR analyses confirmed that hydride ions preferentially occupy oxygen-deficient sites, leading to the formation of dense $BaTiO_{3-x}H_x$ ($BaTiO_{3-x}H_x$). Systematic tuning of the Ti content in the oxygen absorber enabled precise control of hydrogen content, which in turn modulated the electronic transport properties, culminating in a transition from semiconducting to metallic behavior.

Hydrogen desorption studies revealed a distinct dependence on particle size, where surface-near hydrogen desorbed at lower temperatures, while hydrogen embedded deep within crystallites exhibited stronger bonding. Furthermore, deeply incorporated hydride remained in the lattice until the arrival of nitrogen and was released via anion exchange, which enabled a significantly higher level of nitrogen incorporation even under the mild conditions of ambient $N_2$ flow. Notably, the successful formation of single-crystalline oxyhydrides via HPDC using $SrTiO_3$ single crystals marks a significant advancement, unlocking opportunities to explore the intrinsic properties of these metastable phases in the absence of grain boundary effects.

Taken together, these findings establish HPDC as a generalizable platform for accessing dense oxyhydrides with tunable anion composition and transport functionality, thereby paving the way for future investigations into their structural, electronic, and chemical behaviors in both polycrystalline and single-crystalline forms.

## 4. Method

### 4.1 $BaTiO_{3-x}H_x$ and $SrTiO_{3-x}H_x$ from HPDC

The sample space consisted of a vertically stacked structure comprising a hydride source, parent oxide, YSZ (FUJIFILM Wako Pure Chemical Corp.), and an oxygen absorber, as shown in Figure 1(a). High-pressure cells (C & T Factory) with a sample space diameter of φ4.2 mm were modified for HPDC, as reported in a previous study[6a]. A high-pressure apparatus (CTF-MA180P, C & T Factory) was used to apply a pressure of 1 GPa and maintain the temperature at 650 °C for 40 hours. In all synthesis runs, the initial applied voltage was set to 5 V, and the current limit was fixed at 10 mA. *In situ* observation during the HPDC treatment and the effects of voltage application on oxygen extraction are discussed in Figures S6 and S7, respectively.

All parent $BaTiO_3$ pellets were prepared by sintering commercial $BaTiO_3$ (FUJIFILM Wako Pure Chemical Corp.) powder at 1300 °C for 20 hours. For comparison, pellets used in the experiments shown in Figure 3(b), which investigated the effect of initial sintering, were prepared by sintering at 1100 °C or 1200 °C for 20 hours. The oxygen absorber was typically prepared by mixing Ti (UJIFILM Wako Pure Chemical Corp.) and YSZ in a 1:1 molar ratio. The hydrogen concentration in the resulting samples could be systematically tuned by adjusting the Ti content in the oxygen absorber. In the experiments shown in Figure 4, a series of oxygen absorbers with Ti : YSZ molar ratios of 1:30, 1:10, 1:5, 1:2, 1:1.5, and 1:1 were employed. All samples were prepared using $BaTiO_3$ pellets sintered simultaneously at 1300 °C in the same furnace to ensure consistent initial microstructure for Figure 4. $MgH_2$ (UJIFILM Wako Pure Chemical Corp.) was employed as the hydrogen source, owing to its ability to gradually release hydrogen at temperatures above ~600 °C under 1 GPa[7b]. In contrast, in the experiment shown in Figure S2, $CaH_2$ was used instead, as it was considered more suitable than $MgH_2$ for HPDC conducted at 900 °C due to its higher hydrogen release temperature.

$SrTiO_3$ pellets were prepared by sintering commercial $SrTiO_3$ (FUJIFILM Wako Pure Chemical Corp.) powder at 1200 °C for 20 hours. Also, A single-crystalline $SrTiO_3$ (Crystal Base Co., Ltd.) piece, shaped into a cylinder with a diameter of 4.2 mm and a height of 2 mm, was used as the parent material. Typical HPDC conditions used for $BaTiO_3$ were applied for the anion conversion from oxygen to hydride ions.

### 4.2 $BaTiO_{3-x}H_x$ from co-sintering with $CaH_2$

The stoichiometric $BaTiO_3$ powders with 0.1 μm grain size (SAKAI CHEMICAL INDUSTRY Co., LTD.) were preheated at 125 °C for 1 day to remove adsorbed water. Then the dried $BaTiO_3$ powder was mixed with a 3 M excess of $CaH_2$ powders and the mixed powders were pelletized in a glovebox with a dry inert Ar gas (the dew point <−100 °C, the oxygen concentration < 1 ppm). The pressed 10-mmϕ pellet with a thickness of 5 mm was then sealed in an evacuated silica glass ampule (~$10^{-3}$ Pa). The sealed ampule was heated at 450 °C for 5 days. The product was crushed and then washed in a 0.1 M $NH_4Cl$/methanol solution for 1 h in air to remove excess $CaH_2$ and CaO powders, and then dried in an oven at T = 125 °C.

### 4.3 Sample characterization

TPD measurements (MictotracBEL Corp. BELMASS) were performed to estimate the H concentration. The temperature dependence of the sample resistivity was measured using a refrigeration system (Thermal Block Company, SB-1.5KCRS0.5-TRB). The samples, measuring 4.2 x 1.2 x 0.5 $mm^3$, were prepared, and Ag paste was used to contact the gold electrode with the samples. The four-probe method was used to measure the electrical resistivity. The time-of-flight (TOF) powder neutron diffraction was measured in BL-20 at the Material and Life Science Experimental Facility (MLF) of the Japanese Proton Accelerator Research Complex (J-PARC). Rietveld refinements for ND data were performed using the program Z-Rietveld [14]. The in-situ observation of synchrotron X-ray diffraction measurements under high pressure at the BL14B1 experimental station of SPring-8. The chemical state of H was measured by $^{1}$H magic angle spinning (MAS) NMR using a spectrometer (Bruker AVANCE III HD 600WB) at a Larmor frequency of 600.395 MHz. A Bruker MAS probe head (MAS1.3DR) was used with a MAS rotor with 1.3 mm diameter (HZ14752). The sample was packed into the rotor, and the spin rate was 60 kHz. The amount of sample was 0.01326g. The $^{1}$H chemical shifts were referenced to TMS at 0 ppm. Adamantane was used as a second reference material, the central signal of which was set at 1.83 ppm.

**Acknowledgment**

This work was supported by the Japan Science and Technology Agency (JST) CREST (Grant No. JPMJCR19J1), the Japan Society for the Promotion of Science (JSPS) (Grant Nos. 19H02420 and 24K01171). The synchrotron XRD measurements were supported by the QST Advance Characterization "Advanced Research Infrastructure for Materials and Nanotechnology in Japan (ARIM)" (Proposal Nos. JPMXP1222QS0114 and JPMXP1223QS0012) of the Ministry of Education, Culture, Sports, Science and Technology (MEXT). The synchrotron radiation X-ray diffraction experiments were performed using a QST experimental station at the QST beamline BL14B1 and SPring-8, with the approval of the Japan Synchrotron Radiation Research Institute JASRI (Proposal Nos. 2022B3682 and 2023A3682). The neutron diffraction measurements were performed at the beamline BL-20 in the Material and Life Science Experimental Facility (MLF) of the Japanese Proton Accelerator Research Complex (J-PARC) (No.2024A0301). Solid-state NMR measurements were supported by the AIST Nanocharacterization Facility (ANCF) in the Advanced Research Infrastructure for Materials and Nanotechnology (No. JPMXP1224AT5040) sponsored by the Ministry of Education, Culture, Sports, Science, and Technology (MEXT), Japan. We would like to express our gratitude to Ms. Masae Sawamoto, Research Institute for Electronic Science, Hokkaido University, and Ms. Yukino Nishikubo, Innovative Functional Materials Research Institute, National Institute of Advanced Industrial Science and Technology for preparing the HPDC cells and providing insight that greatly assisted the research.